\documentclass[aps,pre,twocolumn,notitlepage,superscriptaddress]{revtex4-2}

\usepackage{amsmath}
\usepackage{amssymb}
\usepackage{hyperref}
\usepackage{graphicx}
\usepackage[arrow, matrix, curve]{xy}
\usepackage{bbm}
\usepackage{bm}
\usepackage{yhmath}
\usepackage{color}
\usepackage{physics}
\usepackage{bbold}
\usepackage{cleveref}
\usepackage{import}
\usepackage{graphicx}
\usepackage{calc}
\graphicspath{{Figures/}}

\usepackage{color}

\usepackage{mathtools}
\usepackage{esint}
\usepackage{upgreek}

\renewcommand\vec[1]{\boldsymbol{\mathrm{#1}}}

\usepackage[usenames,dvipsnames]{xcolor}
\hypersetup{colorlinks=true, linkcolor=BrickRed, urlcolor=blue!50!black, citecolor=blue!50!black}

\begin{document}

	\title{Transition-path sampling for Run-and-Tumble particles}

	\author{Thomas Kiechl}
	\affiliation{Institut f\"ur Theoretische Physik, Universit\"at Innsbruck, Technikerstra{\ss}e 21A, A-6020, Innsbruck, Austria}
	\author{Thomas Franosch}
	\affiliation{Institut f\"ur Theoretische Physik, Universit\"at Innsbruck, Technikerstra{\ss}e 21A, A-6020, Innsbruck, Austria}
	\author{Michele Caraglio}
	\affiliation{Institut f\"ur Theoretische Physik, Universit\"at Innsbruck, Technikerstra{\ss}e 21A, A-6020, Innsbruck, Austria} 
	\email[Electronic mail: ]{michele.caraglio@uibk.ac.at}
	
	\date{\today}
	
	\begin{abstract}
		We elaborate and validate a generalization of the renowned transition-path-sampling algorithm for a paradigmatic model of active particles, namely the Run-and-Tumble particles.
		Notwithstanding the non-equilibrium character of these particles, we show how the consequent lack of the microscopical reversibility property, which is usually required by transition-path sampling, can be circumvented by identifying reasonable backward dynamics with a well-defined path-probability density.
		Our method is then applied to characterize the structure and kinetics of rare transition pathways undergone by Run-and-Tumble particles having to cross a potential barrier in order to find a target.
	\end{abstract}
	
	
	\maketitle

	\section{\label{sec:intro} Introduction}

	Computer simulations play an important role in understanding nature.
	However, several situations are encountered in which the system under investigation displays rare events, i.e., events that occur at times scales several orders of magnitude longer than the typical time scale of the process under consideration.
	Major earthquakes~\cite{Scholz2002,Baiesi2004}, spreading of diseases~\cite{Jeltsch1997}, financial shocks~\cite{Embrechts1997}, and protein folding~\cite{Onuchic1997} are just few possible examples. 
	This kind of events are intuitively challenging standard simulation methods and enhanced sampling techniques have explicitly been developed to overcome this difficulty~\cite{torr1977,Laio2002,dell1998,bolh2002,dell2003,verp2003,verp2005,fara2004,bell2015,alle2009,huss2020}.
	
	Among the various enhanced sampling algorithms, transition-path sampling (TPS)~\cite{dell1998,bolh2002,dell2003}, essentially a Metropolis-Monte-Carlo sampling in the space of reactive trajectories, has the advantages of being relatively simple and providing a completely rigorous sampling of the reactive trajectories between a reactant basin and a target basin.
	In its original formulation, TPS was specifically thought for equilibrium systems, for which it is possible to write down a backward probability, directly related to forward dynamics by microscopic reversibility.
	This includes Brownian dynamics, Monte-Carlo dynamics, and most standard molecular dynamics.
	However, several other processes are intrinsically out-of-equilibrium because they imply time-dependent and/or non-conservative force fields.
	Examples encompass crystal nucleation under shear~\cite{Ackerson1988,Blaak2004}, polymer conformational transitions in hydrodynamic flows~\cite{Schroeder2003}, driven transport through pores~\cite{Henrickson2000}, and active particles~\cite{Elgeti2015,Marchetti2013,bech2016}.
	To address also these cases, other algorithms such as, for instance, the noise-history approach~\cite{Crooks2001}, non-equilibrium umbrella sampling~\cite{Warmflash2007,Dickson2009}, weighted-ensemble method~\cite{Huber1996,Zhang2010}, stochastic-process rare-event sampling~\cite{Berryman2010}, and forward-flux sampling (FFS)~\cite{alle2009,huss2020} have been developed.
	More recently, a promising machine-learning approach based on unsupervised normalizing-flow neural networks was also proposed~\cite{Asghar2024}.
	Though, TPS still presents some strengths that are tough to renounce: namely, besides being analytically understandable and, as already mentioned, exactly harvesting the reactive trajectories, it does not need to know \textit{a priori} a proper reaction coordinate and, in certain situations, it can be more efficient than other methods thanks to its ability to exploit also the backward dynamics.
	
	With these motivations in mind, more recently, two novel methods~\cite{buij2020,Zanovello2021} addressed the challenge of developing a TPS algorithm able to sample rare events in active particle systems.
	The first method, by Buijsman and Bolhuis~\cite{buij2020}, bridges between the original TPS and the FFS methodology:
	As TPS, it performs Monte-Carlo sampling of the trajectory space and is independent of the choice of a reaction coordinate.
	The latter is here replaced by a ``progress variable'' that measures the progress of a trial path along an existing path in the transition path ensemble and is employed to increase the bias toward successful trajectories.
	However, as FFS, this method is partially limited by the impossibility of using backward shooting to speed up the sampling of the reactive path ensemble.
	
	Following a different approach, Zanovello \textit{et al.}~\cite{Zanovello2021} have instead shown that the explicit breaking of microscopic reversibility does not prevent exploiting the original TPS algorithm.
	In fact, in the case of active Brownian particles, they show that, to employ the TPS scheme even when the correct backward dynamics remain elusive, it is enough to prescribe a tentative backward dynamics with a well-defined path probability density and then consistently derive the Metropolis acceptance probability of new reactive trajectories.
	However, even if in principle any tentative backward dynamics would work, efficiency requires that the new trajectories obtained by joining the backward and the forward segments remain ``smooth'', i.e. they look like the trajectories obtained with only forward dynamics.

	The active Brownian particle (ABP)~\cite{bech2016,roma2012} provides a simple stochastic model able to describe with a certain accuracy the properties of motion of a large fraction of existing microswimmers, particularly artificial ones such as the renowned active Janus particles~\cite{Howse2007,Jiang2010,Volpe2011,Kurzthaler2018}.
	In the past decades, it has become the minimal paradigm to investigate irreversible non-equilibrium dynamics and, as such, have been the object of intensive research aiming at understanding both their single-particle and collective behavior in several environments~\cite{Fily2013,Buttinoni2013,Khatami2016,Kurzthaler2016,Malakar2020,Zanovello2021b,Caraglio2022,Steffenoni2017}.
	However, several other examples of active particles exist which are not well represented by the ABP model.
	This is the case of many bacteria whose self-propulsive motion is characterized by a sequence of linear straight ``\textit{runs}'' at almost constant speed punctuated by reorientation events (also called tumbling events or ``\textit{tumbles}'') occurring stochastically.
	This Run-and-Tumble dynamics were first introduced to describe the motion of \textit{Escherichia coli} bacteria~\cite{Berg2004} and since then it has also been thoroughly investigated~\cite{Schnitzer1993,Hill2007,Tailleur2008,Tailleur2009,Angelani2009,Martens2012,Kurzthaler2024}.
	
	In this paper we show how the TPS algorithm can also be adapted to a Run-and-Tumble particle (RTP), thus extending the range of applicability of this powerful enhanced sampling technique to virtually all the relevant active particle models.
	Our findings are then exploited to investigate how Run-and-Tumble particles solve a target-search problem which entails going beyond some energetic barrier.
	Similar issues recurs in many realistic scenarios that active agents have to face~\cite{volp2017,beni2011,visw1999,beni2005,beni2006,visw2008}, with a typical example given by bacteria foraging for nourishment in environments with scarce resources.
	In particular, the knowledge of the transition-path ensemble paved by TPS allows us to characterize the Run-and-Tumble particle target-search behavior in terms of relevant quantities, such as the transition-path time (TPT) distribution~\cite{neup2016,sega2007,cara2018} and the transition probability density with its associated reactive current~\cite{wein2010,Metzner2006,Bartolucci2018}, which also allows identifying the navigation patterns.

	\section{\label{sec:methods} Methods}
	
	\subsection{\label{sec:model}Model}
	We consider an RTP navigating in a two-dimensional energy landscape, starting from an initial region R (also named the reactant region in the remainder of the text) and having to overcome an energy barrier in order to reach a target region T.
	The Run-and-Tumble motion of the microswimmer comprises two distinct phases:
	A passive phase, during which the motion of the particle is governed by standard translational diffusion with diffusion coefficient $D$, and an active phase, during which the particle is also able to self-propel with constant velocity $v$ in a given direction $\vec{u}=(\cos \vartheta,\sin \vartheta)$.
	The directional angle $\vartheta$ is also undergoing a diffusion process with rotational diffusion coefficient $D_{\rm rot}$.
	However, upon transitioning from the passive to the active phase, the particle undergoes a reorientation event, termed tumbling, leading to complete randomization of its orientation.
	In the following, the phase of the particle is encoded in a binary variable $s$ being equal to $0$ if the phase is passive and $1$ if the phase is active.
	As a starting point for a suitable TPS formulation, we here consider a discrete dynamics, with integretion time step $\Delta t$.
	Thus, at time $t=i\Delta t$, the state of the particle is completely characterized by a tuple $\omega_i := (s_i,\vec{r}_i,\vartheta_i)$, with $\vec{r}_i:=(x_i,y_i)$ its position.
	We also assume that the dynamics is Markovian, meaning that the probability of the state visited at step $i+1$ depends only on the state at step $i$ and not on the states visited in earlier steps.
	Finally, we also assume that both phases have exponentially distributed durations $\propto e^{- \lambda t}$ where $\lambda$ is a rate depending on the specific phase.
	In particular we distinguish between a passive-to-active rate $\lambda_{0\to1}$ and an active-to-passive rate $\lambda_{1\to 0}$.
	
	The equations of motion, discretized according to the It\^{o} rule, of the RTP in a conservative energy landscape $U(\vec{r})$ then reads
	\begin{align}
		s_{i+1}  & \!=\!
			\begin{cases}
				s_{i}  & \!\!\! \text{ with prob. } p_{i} := e^{-\lambda_{s_{i} \to (1 - s_{i})} \Delta t} \, ,\\
				1-s_{i}  & \!\!\!  \text{ with prob. } 1-p_{i} \, ,\\
			\end{cases} \label{eq:transitionprob_forward} \\
		\vec{r}_{i+1} & \!=\! \vec{r}_{i} - \mu \vec{\nabla}U(\vec{r}_{i}) \Delta t+ \sqrt{2 D \Delta t } \, \vec{\xi}_{i}+  v  s_{i} \vec{u}_{i} \Delta t \, , \label{eq:eomrtpt_trans} \\
		{\vartheta}_{i+1} & \!=\! \begin{cases}
			2\pi\, \zeta_i                                     & \text{if } s_{i} = 0 \land s_{i+1}=1 \,  \\
			\vartheta_{i} + \sqrt{2 D_{\rm rot} \Delta t} \, \eta_{i} & \text{else}
		\end{cases} \label{eq:eomrtpt_vartheta}
	\end{align}
	Here, the effective mobility is denoted by $\mu$ and the components of the vector noise $\boldsymbol{\xi}_{i} = (\xi_{x,i}, \xi_{y,i})$ and the scalar noise $\eta_{i}$ are independent random variables, distributed according to a normal distribution with zero mean and unit variance.
	Finally, $\zeta_{i}$ is a random variable distributed according to a uniform distribution in the interval $[0,1]$.
	Note that a tumble event, leading to a completely random reorientation of the self-propulsion direction, occurs only at the beginning of a new active phase after a passive phase has ended.
	Note also that for passive-to-active rate approaching infinity ($\lambda_{0 \to 1} \to \infty$), our model in the continuous limit (see Appendix~\ref{sec:fokkerplanck}) reduces to a run-and-tumble model in which ABP phases are separated by instantaneous tumbling events (see Appendix~\ref{appendix_infinityrate}).
	
	We benchmark the performance of the TPS algorithm adapted to RTPs in a paradigmatic two-well potential landscape provided by
	\begin{align}\label{eq:potential}
		U(x,y)=k_x(x^2-x_0^2)^2+\frac{k_y}{2}y^2 \, ,
	\end{align}
	where $\pm x_0$ specifies the positions of the two potential minima, $k_x x_0^4$ corresponds to the height of the energy barrier along the axis $y=0$, and the ratio $k_y/k_x x_0^2$ characterizes the shape of the landscape.
		\begin{figure}[h]
		\centering
			\includegraphics[width=1.0\columnwidth]{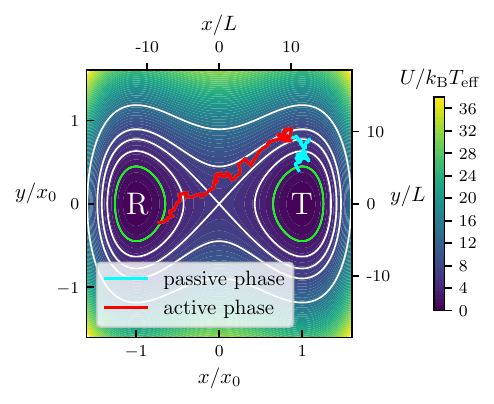}
			\caption{Potential landscape as given in Eq.~\eqref{eq:potential} for barrier height $k_x x_0^4 = 6k_BT$ and ratio $k_y/k_x x_0^2 = 3.3$. 
			The reactant R and the target T basins are encircled in green. 
			A reactive RTP trajectory obtained by integrating Eqs.~(\ref{eq:transitionprob_forward}-\ref{eq:eomrtpt_vartheta}) is depicted within the potential landscape, with active phase marked in red and passive phase in blue. \label{fig:1} 			
			}
	\end{figure}
	The energy landscape is displayed in Fig.~\ref{fig:1} together with an example of an RTP trajectory connecting the reactant to the target. 
	Although the system under consideration is intrinsically out of equilibrium, energy can be expressed in units of the thermal energy $k_\text{B}T_\text{eff}:=D/\mu$, where $T_\text{eff}$ is an effective temperature corresponding to the bath temperature for a completely passive particle.
	We define the length unit $L:= k_\text{B}T_\text{eff}/F_{\text{max}}$ starting from the the maximal force $F_{\text{max}}$ along the minimum energy path connecting R and T.
	Thus, $L$ can be interpreted as the typical length that a passive particle can explore due to thermal fluctuation, when moving in a v-shaped potential with slope $F_{\text{max}}$.
	This maximal force is found at $(\pm x_0/\sqrt{3}, 0)$ and evaluates to  $F_{\text{max}}=8k_xx_0^3/\sqrt{27}$.
	Finally, the time unit is fixed as $\tau:= L^2/D$.
	
	The activity level and the self-propulsion direction's angular diffusion, characterizing the active phase in our RTP model, are encoded in two dimensionless parameters: The P\'eclet number $\text{Pe}:= vL/D$, which quantifies the importance of the self-propulsion with respect to the diffusive motion, and the persistence $\ell^*:= v/D_{\rm rot} L$, characterizing the typical distance over which the trajectories of an active particle remain relatively straight in units of $L$.
	
	Here, we define the reactant region as $\text{R} = \left\{\omega | U(\vec{r}) \leq 2k_\text{B}T_\text{eff} \land x<0\right\}$ and the target region $\text{T} = \left\{\omega | U(\vec{r}) \leq 2k_\text{B}T_\text{eff} \land x>0\right\}$.
	A trajectory $\mathcal{W}=(\omega_0, \omega_1, \dots ,\omega_N)$ is called reactive if it starts in the reactant region ($\omega_0 \in \text{R}$), ends in the target ($\omega_N \in \text{T}$), and all intermediate states belong to the transition region $(\omega_{i} \notin \text{T} \cup \text{R}, i=1,\ldots,N-1$). 
	The time duration of a given reactive path $(\omega_0, \omega_1, \dots ,\omega_N)$, $t_\text{TPT}=N\Delta t $, is called its transition-path time.
	Its average value $\langle t_\text{TPT} \rangle$ and distribution $P(t_\text{TPT})$, as obtained from several reactive paths, are the main observables here exploited to test the validity of the TPS algorithm adapted to RTPs.

	\subsection{\label{sec:TPS}Transition Path Sampling for Run-and-Tumble particles}
	
\begin{figure}[t!] 
		\centering
		\includegraphics[width=1.0\columnwidth]{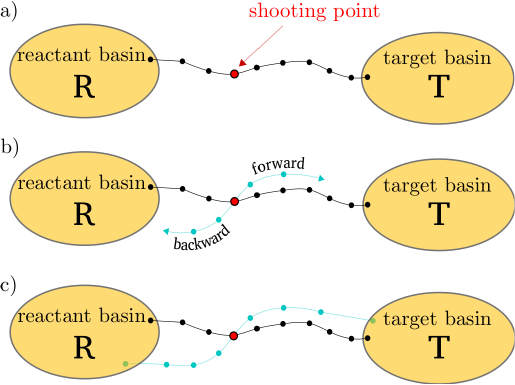}
		\caption{Illustration of the TPS algorithm. 
			\textbf{(a)} Choose with uniform probability a shooting point on the old path connecting the reactant basin R to the target T;
			\textbf{(b)} Evolve the shooting point forward and backward in time until the two endpoints reach either R or T;
			\textbf{(c)} If the new path is reactive, accept or reject it according to Eq.~\eqref{eq:Pacc_general}.
		 \label{fig:2}}	
	\end{figure}

	The key idea of the Transition-Path Sampling (TPS) algorithm lies in employing a Monte-Carlo sampling approach to efficiently generate a Markov chain of reactive paths starting from an arbitrary initial reactive path, see Fig.~\ref{fig:2} for an illustration.
	After being generated from an old trajectory $\mathcal{W}^{\rm old}$, a new trajectory $\mathcal{W}^\text{new}$ is then accepted or rejected according to the Metropolis acceptance probability
	\begin{align}\label{eq:Pacc_general}
		\mathcal{P}_\text{acc}&[\mathcal{W}^{\rm old}\to \mathcal{W}^\text{new}] = h[\mathcal{W}^\text{new}]
		\notag\\ &\times \min\left\{1, \frac{\mathcal{P}[\mathcal{W}^\text{new}] \mathcal{P}_\text{gen}[\mathcal{W}^\text{new} \to \mathcal{W}^{\rm old}]}
		{\mathcal{P}[\mathcal{W}^{\rm old}] \mathcal{P}_\text{gen}[\mathcal{W}^{\rm old} \to \mathcal{W}^\text{new}]} \right\},
	\end{align}
	which derives from the imposed detailed-balance condition in the space of reactive trajectories.
	
	Here, the characteristic function $h\left[\mathcal{W}\right]$ equals zero if $\mathcal{W}$ is not a reactive path and one otherwise, which ensures that only new paths that are reactive will be accepted.
	
	The probability of path $\mathcal{W}$ reads 
	\begin{align}\label{eq:P(W)_general}
		\mathcal{P}[\mathcal{W}]=\rho(\omega_0) \prod_{i=0}^{N-1}\mathbb{P}_{\! \Delta t}(\omega_{i+1}| \omega_{i}),
	\end{align}
	where $\rho(\omega_0)$ is the steady-state probability of being in the initial state $\omega_0$, and $\mathbb{P}_{\! \Delta t}(\omega_{i+1}| \omega_{i})$ is the probability of transitioning from the state $\omega_{i}$ to the state $\omega_{i+1}$during a time step $\Delta t$.
	The latter can be directly derived from the equations of motion, Eqs.~(\ref{eq:transitionprob_forward}-\ref{eq:eomrtpt_vartheta}),
	\begin{align}\label{eq:p_singlestep_forward}
		\mathbb{P}_{\! \Delta t}&(\omega_{i+1} |\omega_{i})= p(s_{i+1} | s_{i}) \nonumber \\
		& \!\!\! \times \! \dfrac{1}{4\pi D\Delta t}
			\exp \! \left( - \dfrac{\left( \vec{r}_{i+1} \!-\! \vec{r}_{i} \!+\! \mu \vec{\nabla} U (\vec{r}_{i}) \Delta t \!-\! v s_{i} \vec{u}_{i} \Delta t \right)^2}{4 D \Delta t} \right) \nonumber \\ 
		& \!\!\! \times \! \left\lbrace 
		\begin{array}{l}
			\dfrac{1}{2\pi} \qquad \qquad \qquad \text{if } s_{i}=0 \land s_{i+1}=1 \, ,
			\\
			\\
			\dfrac{1}{\sqrt{4\pi D_{\rm rot} \Delta t}} \exp  \left(-\dfrac{( {\vartheta}_{i+1} \!-\! \vartheta_{i})^2}{4 D_{\rm rot} \Delta t} \right) \qquad \text{else} \, .
		\end{array} \right.  
	\end{align}
	Here, the factor in the first line, $p(s_{i+1} | s_{i})$, refers to the probability of going from the phase $s_{i}$ to phase $s_{i+1}$ as encoded in Eq.~\eqref{eq:transitionprob_forward}
	\begin{align} \label{eq:phase_transition_prob}
		p(s_{i+1} | s_{i}) := 
		\begin{cases}
			p_{i} & \mbox{if } s_{i+1} = s_{i} \, , \\
			1-p_{i} & \mbox{else} \, ,
		\end{cases}
	\end{align}
	while the second and third lines derive from Eq.~\eqref{eq:eomrtpt_trans} and~\eqref{eq:eomrtpt_vartheta}, respectively.
	
	Finally, $\mathcal{P}_\text{gen}[\mathcal{W} \to \mathcal{W}']$ is the probability of generating the path $\mathcal{W}'$ starting from $\mathcal{W}$ according to
	\begin{align}\label{eq:P_gen_general}
		\mathcal{P}_{\rm gen} & \left[\mathcal{W} \to \mathcal{W}' \right] = 
		    \mathcal{P}_{\rm sel} \left(\omega_m | \mathcal{W} \right) \mathcal{P}_{\rm pert} (\omega_m \to \omega_n' ) \notag \\
		& \times \prod_{i=n}^{N'-1} \mathbb{P}_{\! \Delta t}(\omega_{i+1}'|\omega_{i}')\prod_{i=0}^{n-1} \bar{\mathbb{P}}_{\! \Delta t}(\omega_{i}'|\omega_{i+1}') \, ,
	\end{align}
	where $\mathcal{P}_{\rm sel}(\omega_m|\mathcal{W}) = 1/N$ is the probability of uniformly selecting the state $\omega_m$ from the path $\mathcal{W}$ having length $N$.
	With the aim of enhancing the acceptance ratio, a non-uniform selection probability of the shooting point from the path $\mathcal{W}$ may also be adopted but this goes beyond the scope of the present manuscript.
	Further, $\mathcal{P}_{\rm pert}\left(\omega_m \to \omega_n' \right)$ is the probability of perturbing the state $\omega_m$ into a new state $\omega_n'$, which in the following will be referred as the \textit{shooting point}.
	Due to the stochastic nature of its dynamics, for an RTP the perturbation of the shooting point is not necessary~\cite{dell2003} and we then set $\omega_n'=\omega_m$ with probability equal to one.
	The first product in the second line of Eq.~\eqref{eq:P_gen_general} represents the probability of generating a segment of the new path $\mathcal{W}'$ connecting the shooting point $\omega_n'$ to the state $\omega_{N'}' \in$ T by integrating the equations of motion~(\ref{eq:transitionprob_forward}-\ref{eq:eomrtpt_vartheta}) (\textit{forward shooting}).
	Similarly, the second product corresponds to the probability of generating a segment connecting the shooting point $\omega_n'$ to the state $\omega_{0}' \in$ R by evolving the dynamics backward in time (\textit{backward shooting}).
	Note that, Eq.~\eqref{eq:P_gen_general} represents only the probability of generating a given path $\mathcal{W}'$ starting from another path $\mathcal{W}$ while the acceptance rate is then determined by the acceptance probability, as reported in Eq.~\eqref{eq:Pacc_general}.
In particular, Eq.~(9) does not require that $\mathcal{W}'$ and $\mathcal{W}$ are reactive paths but, because of the characteristic function $h[\mathcal{W}]$ appearing in Eq.~\eqref{eq:Pacc_general}, new paths that are not reactive will be rejected with probability $1$.
On the other hand, new paths that are reactive have a finite probability  $\mathcal{P}_{\rm acc}$ (with $0 < \mathcal{P}_{\rm acc} \leq 1$) of being accepted.
However, practically, from the shooting point one integrates the forward and the backward equations of motion until both the pieces of the arising path reach one of the basins and this choice directly prevents the generation of trajectories having one of the endpoints in the transition region (i.e. not in $T \cup R$).
	
	The single-step contributions to the forward shooting are given by discrete forward propagators, $\mathbb{P}_{\! \Delta t}$, as specified in Eq.~\eqref{eq:p_singlestep_forward}, those in the backward shooting are discrete backward propagators, $\bar{\mathbb{P}}_{\! \Delta t}$, that, for an active particle, are not univocally defined.
	In fact, the lack of microscopic reversibility makes the dynamics of an active particle intrinsically irreversible and there is no consensus on which backward dynamics should be assumed~\cite{Fodor2022}.
	However, as demonstrated in Ref.~\cite{Zanovello2021}, in principle, one could formulate any set of backward equations of motion as long as this is accounted for in the computation of the acceptance probability, Eq~\eqref{eq:Pacc_general}.
	Nonetheless, the efficiency of sampling new reactive paths (by joining the forward and the backward branches departing from the shooting point) increases if we make an educated guess for the backward dynamics that produces paths that are similar to those generated by following the forward dynamics.

	Here, we update a state backward in time according to
	\begin{align}
		s_{i}  & \!=\!
			\begin{cases}
				s_{i+1}  & \!\!\! \text{ with prob. } p_{i+1} \, ,\\
				1-s_{i+1}  & \!\!\!  \text{ with prob. } 1-p_{i+1} \, ,\\
			\end{cases} \label{eq:transitionprob_backward} \\
		{\vartheta}_{i} & \!=\! \begin{cases}
			2\pi\, \zeta_{i+1}                                      & \text{if } s_{i} \!=\! 0 \land s_{i+1} \!=\! 1 \, ,  \\
			\vartheta_{i+1} + \sqrt{2 D_{\rm rot} \Delta t} \, \eta_{i+1} & \text{else} \, ,
		\end{cases} \label{eq:eomrtpt_vartheta_back} \\
		\vec{r}_{i} & \!=\! \vec{r}_{i+1} \!-\! \mu \vec{\nabla}U(\vec{r}_{i+1}) \Delta t \!+\! \sqrt{2 D \Delta t } \, \vec{\xi}_{i+1} \!-\!  v  s_{i} \vec{u}_{i} \Delta t \, . \label{eq:eomrtpt_trans_back}
	\end{align}
	Note that, in our definition of the backward dynamics, the self-propulsion direction is first updated and then used when updating the position of the particle.
	Furthermore, we stress the fact that, in Eq.~\eqref{eq:eomrtpt_trans_back}, the self-propulsion term is treated as odd under time-reversal symmetry while the force field is treated as even with respect to the same symmetry~\cite{Fodor2022}.
	These choices ensure that the trajectories sampled in the TPS scheme resemble those obtained by considering only the forward dynamics and that in the passive limit ($\text{Pe}=0$) the forward and backward probabilities are indistinguishable~\cite{dell2003}.
	
	The discretized backward dynamics readily results in a discrete single-step backward propagator given as
	\begin{align}\label{eq:p_singlestep_backward}
		&\bar{\mathbb{P}}_{\! \Delta t}(\omega_{i} |\omega_{i+1})= p(s_{i} | s_{i+1}) \nonumber \\
		& \; \times \! \dfrac{1}{4\pi D\Delta t}
			\exp \! \left( - \dfrac{\left( \boldsymbol{r}_{i} \!-\! \vec{r}_{i+1} \!+\! \mu \boldsymbol{\nabla} U (\vec{r}_{i+1}\!) \Delta t \!+\! vs_{i}\boldsymbol{u}_{i}\Delta t \right)^2}{4 D \Delta t} \right)\nonumber \\ 
		& \; \times \! \left\lbrace 
		\begin{array}{l}
			\dfrac{1}{2\pi} \qquad \qquad \qquad \text{if } s_{i}=0 \land s_{i+1}=1 \, ,
			\\
			\\
			\dfrac{1}{\sqrt{4\pi D_{\rm rot} \Delta t}} \exp  \left(-\dfrac{( {\vartheta}_{i} \!-\! \vartheta_{i+1})^2}{4 D_{\rm rot} \Delta t} \right) \qquad \text{else} \, .
		\end{array} \right.  
	\end{align}
	
	To find an explicit expression for the acceptance probability, Eq.~\eqref{eq:Pacc_general}, it is convenient to rewrite, in the single-step forward propagators, the terms attaining to the update of the position as
	\begin{align}
		& \dfrac{1}{4\pi D\Delta t} \exp \! \left( - \dfrac{\left( \vec{r}_{i+1} \!-\! \vec{r}_{i} \!+\! \mu \vec{\nabla} U (\vec{r}_{i}) \Delta t \!-\! v s_{i} \vec{u}_{i} \Delta t \right)^2}{4 D \Delta t} \right) \nonumber \\
		& \quad = \pi(\vec{r}_{i+1}|\vec{r}_{i}) \exp \left(- \dfrac{v^2 s_{i}^2 \Delta t}{4D}  \right)  \nonumber \\
		& \qquad \times \exp \! \left( \dfrac{v s_{i}}{2D} \vec{u}_{i} \cdot \left( \vec{r}_{i+1} \!-\! \vec{r}_{i} \!+\! \mu \vec{\nabla} U (\vec{r}_{i}) \Delta t \right) \right) \, ,
	\end{align}
	where we have factorized the probability $\pi(\vec{r}_{i+1}|\vec{r}_{i})$ for a passive particle going from $\vec{r}_{i}$ to $\vec{r}_{i+1}$ in a time step $\Delta t$.
	Doing the same for the single-step backward propagators, we have
	\begin{align}
		& \dfrac{1}{4\pi D\Delta t} \exp \! \left( - \dfrac{\left( \vec{r}_{i} \!-\! \vec{r}_{i+1} \!+\! \mu \vec{\nabla} U (\vec{r}_{i+1}) \Delta t \!+\! v s_{i} \vec{u}_{i} \Delta t \right)^2}{4 D \Delta t} \right) \nonumber \\
		& \quad = \pi(\vec{r}_{i+1}|\vec{r}_{i}) \dfrac{\pi(\vec{r}_{i})}{\pi(\vec{r}_{i+1})} \exp \left( - \dfrac{v^2 s_{i}^2 \Delta t}{4D}  \right)  \nonumber \\
		& \qquad \times \exp \! \left( \dfrac{v s_{i}}{2D} \vec{u}_{i} \cdot \left( \vec{r}_{i+1} \!-\! \vec{r}_{i} \!-\! \mu \vec{\nabla} U (\vec{r}_{i+1}) \Delta t \right) \right) \, ,
	\end{align}
	where we have also applied the detailed-balance condition holding for a passive particle
	\begin{align}
		\pi(\vec{r}_{i+1}) \pi(\vec{r}_{i}|\vec{r}_{i+1}) = \pi(\vec{r}_{i}) \pi(\vec{r}_{i+1}|\vec{r}_{i}) \, ,
	\end{align}	
	with $\pi(\vec{r}) \propto e^{-U(\vec{r})/k_B T_{\rm eff}}$ the Boltzmann distribution for a passive particle.

	Combining all terms, Eq.~\eqref{eq:Pacc_general} is evaluated to 
	\begin{align} \label{eq:acceptance_probability_RTPT}
		& \mathcal{P}_\text{acc}[\mathcal{W}^{\rm old}\to \mathcal{W}^\text{new}] = h[\mathcal{W}^\text{new}] \min \Bigg\{ \! 1,  \dfrac{N^{\rm old}}{N^{\rm new}}   \nonumber \\
		& \quad \times   \dfrac{\rho(\omega_0^{\rm new})}{\rho(\omega_0^{\rm old})}
		                 \dfrac{\pi(\vec{r}_0^{\rm old})}{\pi(\vec{r}_0^{\rm new})}  
		                  \dfrac{1-p_{0}^{\rm new}}{1-p_{0}^{\rm old}}\nonumber \\
		& \quad \times \prod _{i=0}^{m-1} \exp \!\! \left( -\dfrac{v \mu \Delta t}{2D} s_{i}^{\rm old} \vec{u}_{i}^{\rm old} \! \cdot \! \big( \vec{\nabla}U(\vec{r}_{i}^{\rm old}) \!+\! \vec{\nabla}U(\vec{r}_{i+1}^{\rm old})    \big) \!   \right)   \nonumber \\
		& \quad \times \prod _{i=0}^{n-1} \exp \!\! \left( \dfrac{v \mu \Delta t}{2D} s_{i}^{\rm new} \vec{u}_{i}^{\rm new} \! \cdot \! \big( \vec{\nabla}U(\vec{r}_{i}^{\rm new}) \!+\! \vec{\nabla}U(\vec{r}_{i+1}^{\rm new})    \big) \!   \right) \!\! \Bigg\} \, ,
	\end{align}
	where we used the fact that 
		\begin{align}
		\prod_{i=0}^{\ell-1} \dfrac{ p(s_{i+1} | s_{i}) }{ p(s_{i} | s_{i+1}) } =  \dfrac{1-p_{0}}{1-p_{\ell}} \, ,
	\end{align}
	and that the shooting point is not perturbed ($s_n^{\rm new} = s_m^{\rm old}$).
	In the limit of vanishing activity ($v \to 0$) we recover the standard TPS formula for passive Brownian dynamics~\cite{dell1998}.
	We also recover such a formula if we consider a purely passive initial reactive path and a vanishing passive-to-active rate ($\lambda_{0\to1} \to 0$).
	On the other hand, if the initial reactive path is purely active and the active-to-passive rate vanishes ($\lambda_{1\to 0} \to 0$), we recover the acceptance probability of an ABP particle as reported in Ref.~\cite{Zanovello2021}.
 	A last remark is in order: The discussed algorithm works provided that a path integral, or equivalently a Fokker-Planck equation can be formulated, which we both elaborate in Appendix~\ref{sec:fokkerplanck} and~\ref{sec:pathint}.

	\section{\label{sec:results} Results and discussion}
	
	In the present work, we are mainly focused on how the results are changing when varying the passive-to-active rate $\lambda_{0\to1}$ and the active-to-passive rate $\lambda_{1\to 0}$.
	Thus, throughout this manuscript, we keep the barrier height $k_x x_0^4 = 6k_BT$ and ratio $k_y/k_x x_0^2 = 3.3$ fixed.
	We also fix the free parameters characterizing the behavior of the active phase to $\text{Pe}=0.39$ and $\ell^*= 33.7$.
	For these parameters, the behavior of the RTP in the active phase resembles that of ABP with large activity, as detailed in Ref.~\cite{Zanovello2021b} where also other different values of the P\'eclet number and of the persistence are investigated.
	
	Considering the paradigmatic double-well potential described in Section~\ref{sec:model}, we use the TPS algorithm previously described to characterize the transition-path times statistics.
	This observable has received increasing attention~\cite{neup2016,chun2009,sega2007,zhan2007,cara2018,carl2018,cara2020} since it carries important information about the reactive dynamics.
	To validate our algorithm, we also benchmark the obtained results against those obtained by directly integrating the equations of motion~(\ref{eq:transitionprob_forward}-\ref{eq:eomrtpt_vartheta}).
	
	Before we apply the TPS algorithm, note that the steady-state distribution in the reactant basin is needed to compute the acceptance probability for new reactive paths, see Eq.~\eqref{eq:acceptance_probability_RTPT}.
	However, an analytical expression for the steady-state distribution is generally unknown for active particles~\cite{Caraglio2022}.
	Therefore, we first numerically estimate $\rho(\omega)$ as the relative occupation frequency of microstates in the reactant basin $\omega \in \text{R}$ by performing standard Langevin simulations, i.e. by repeatedly integrating the equations of motion, Eqs.~(\ref{eq:transitionprob_forward}-\ref{eq:eomrtpt_vartheta}).
	
	\begin{figure}[t!]
		\centering
		\includegraphics[width=1.0\columnwidth]{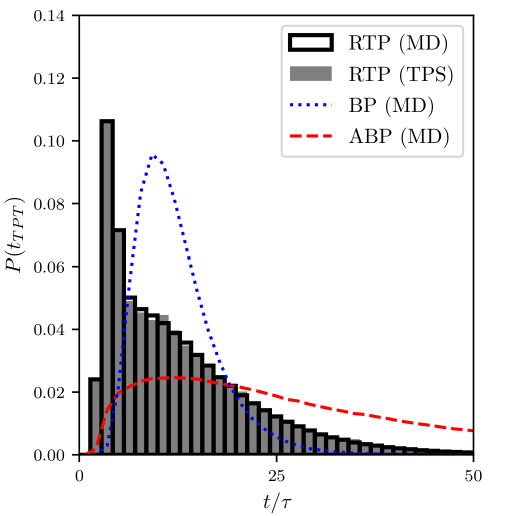}
		\caption{Transition-Path-Time distribution $P(t_{\rm TPT})$ for Run-and-Tumble particles (RTPs) for $\lambda_{1\to0}= \lambda_{0\to1} =0.12/\tau$ generated with the Transition-Path Sampling (TPS) (filled bars) and compared to that obtained from standard MD simulations (black empty bars). For comparison we also report the TPT distribution of a fully passive Brownian particle -BP- (dotted cyan line) and that of an ABP having the same $\text{Pe}$ and $\ell^*$ as the RTP (dashed red line). Statistics collected from $10^6$ reactive paths for standard MD simulations and $10^8$ reactive paths for TPS. \label{fig:3}}
	\end{figure}

	The TPT distribution $P(t_\text{TPT})$ obtained using the TPS algorithm closely matches the one sampled by direct integration of Eqs.~(\ref{eq:transitionprob_forward}-\ref{eq:eomrtpt_vartheta}), as exemplary displayed in Fig.~\ref{fig:3} where such a distribution is reported for rates $\lambda_{0\to1} = \lambda_{1\to 0} =0.12/\tau$.
	In the example displayed in Fig.~\ref{fig:3}, we recognize some typical features of the TPT distribution obtained for ABPs~\cite{Zanovello2021,Zanovello2021b}.
	In particular, we point out that the distribution of an RTP peaks at shorter times but has a longer tail with respect to the TPT distribution shown by a purely passive particle.
	However, the tail for the RTP decays quicker than that of an ABP having the same P\'eclet ($\text{Pe}=0.39$) and persistence ($\ell^*= 33.7$) of the RPT in its active phase.
	We will argue about the mechanisms leading to this behavior later when discussing the reactive probability density and the reactive current.
	
	We have checked that the ensembles of trajectories obtained by exploiting the two methods yield the same $P(t_\text{TPT})$ also across all combinations of rates $\lambda_{0\to 1}, \lambda_{1\to 0} \in [0, 10/\tau]$.
	Note that a rate equal to $10/\tau$ is large enough to allow several phase changes even during a relatively short reactive trajectory.
	Note also that the cases $\lambda_{0 \to 1} =0$ and $\lambda_{1 \to 0} = 0$ correspond to a fully passive and a fully active behavior, respectively.
	Only the case $\lambda_{1 \to 0} = \lambda_{0 \to 1} =0$ is excluded from our analysis because it completely depends on the very first phase chosen for the construction of the first initial arbitrary reactive path.

	\begin{figure}[t]
		\centering
		\includegraphics[width=1.0\columnwidth]{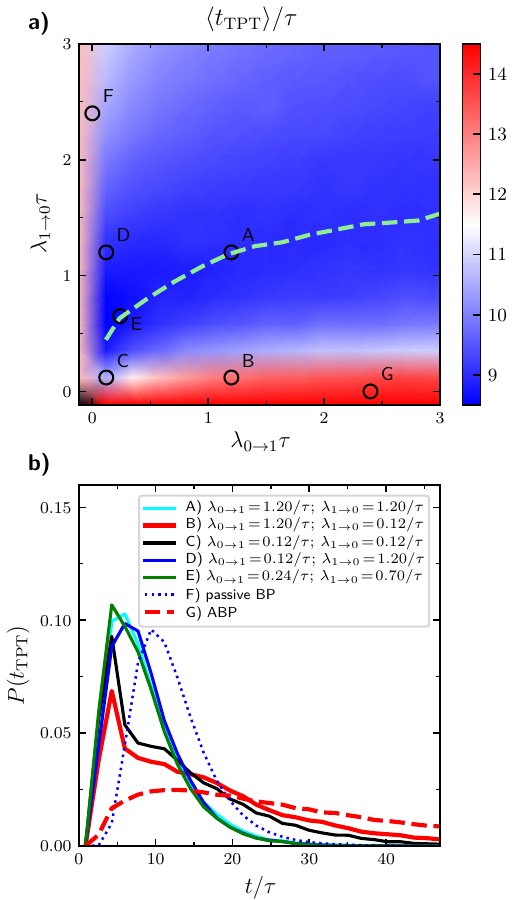}
		\caption{\textbf{(a)} Average TPT, $\langle t_\text{TPT} \rangle$, as a function of the passive-to-active rate $\lambda_{0\to1}$ and of the active-to-passive rate $\lambda_{1\to0}$, as obtained by exploiting the TPS algorithm to collect $10^8$ reactive paths for each combination of $\lambda_{0\to1}$ and $\lambda_{1\to 0}$. The green dashed line indicates the value of $\lambda_{1\to 0}$ as a function of $\lambda_{0\to1}$ for which the average TPT is minimal. \textbf{(b)} TPT distribution for the combination of $\lambda_{0\to1}$ and $\lambda_{1\to 0}$ highlighed by the circles A-G in panel (a). See Appendix~\ref{sec:appD} for a comparison with the distributions obtained from standard MD simulations. \label{fig:4}
		}
	\end{figure}	
	
	In Fig.~\ref{fig:4}, we systematically investigate the average TPT behavior of RTPs as a function of $\lambda_{0\to1}$ and $\lambda_{1\to0}$ as obtained by exploiting the TPS algorithm.
	For purely passive ($\lambda_{0\to1}=0$) and active ($\lambda_{1\to0}=0$) systems, a constant $\langle t_\text{TPT} \rangle$ is observed independently of the respective other rate.
	In particular, for purely passive particles we find an average transition time of about $12.3\tau$ which is shorter than that of about $33.18\tau$ for the ABP particle.
	Interestingly, the RTPs display a large range of rate pairs ($\lambda_{0\to1},\lambda_{1\to0}$) for which the mean transition-path time is even smaller than that of a passive particle, with a minimum of about $8.5\tau$ for $\lambda_{0\to1}= 0.2/\tau$ and $ \lambda_{1\to 0}= 0.7/\tau$ (point E in Fig~\ref{fig:4}a).
	In Fig.~\ref{fig:4}, we also report with a dashed green line the value of $\lambda_{1\to 0}$ which minimizes the average TPT for a given value of $\lambda_{0\to 1}$.
	Such a \textit{valley} of minima is very broad and when departing from it going towards larger values of $\lambda_{1\to 0}$ or smaller $\lambda_{0\to 1}$ the RTP behaves more and more as a passive one, while when moving in the other direction the RTP behaves more and more as an ABP.
	The TPT distribution for rate combinations such that the average TPT is close to the minimal value (point A, D, and E in Fig.~\ref{fig:4}a) looks like the distribution obtained for a passive particle (point F in Fig.~\ref{fig:4}a).
	However, they are shifted towards shorter times, see Fig.~\ref{fig:4}b.
	On the other hand, as already noticed in Fig.~\ref{fig:3}, when the active-to-passive rate $\lambda_{1\to 0}$ is small (point B, and C in Fig.~\ref{fig:4}a), the TPT distribution develops a peak at short times and a longer tail but remains quite different from the distribution of an ABP, see Fig.~\ref{fig:4}b.

\begin{figure*}[t!] 
	\centering
	\includegraphics[width=1.0\textwidth]{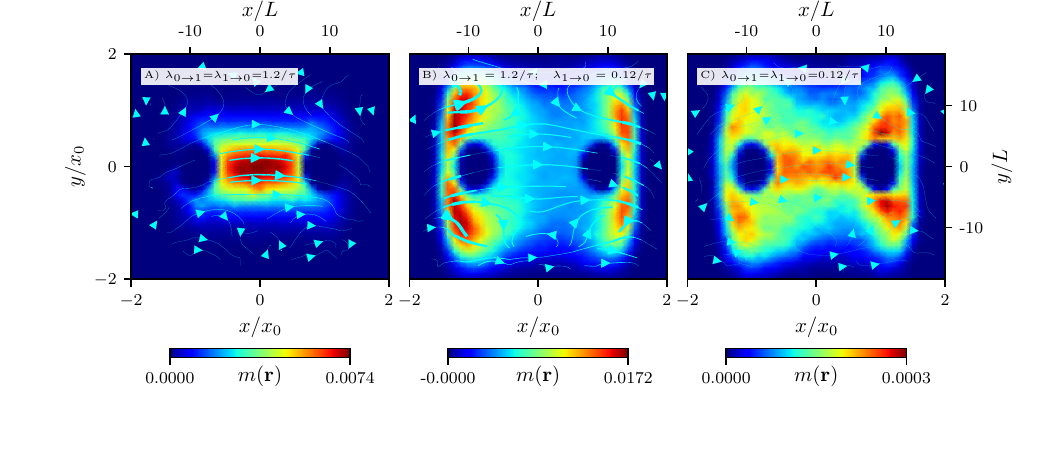}
	\caption{Reactive probability density $m(\vec{r})$ (colormap) and reactive current $\vec{J}(\vec{r})$ (cyan field lines) for an RTP with different combinations of the active-to-passive rate and passive-to-active rate.\label{fig:5}}	
\end{figure*}

	To investigate the origin of this phenomenology, it is instructive to analyze the transition-path density $m(\vec{r})$ and the transition current $\vec{J}(\vec{r})$ as a function of the position $\vec{r}$.
	Here, $m(\vec{r})$ measures the probability that a transition path visits a specific position $\vec{r}$ in the reactive region, while $\vec{J}(\vec{r})$ provides the information on the probability current.
	When the passive-to-active rate $\lambda_{0\to1}$ and the active-to-passive rate $\lambda_{1\to0}$ are such that the average TPT is small, $m(\vec{r})$ and $\vec{J}(\vec{r})$ behave similarly to the case of a completely passive particle (compare Fig.~\ref{fig:5}a to Fig.~3a in Ref.~\cite{Zanovello2021}) with reactive paths mainly exploring only a small region around the minimal energy path connecting R to T.
	Essentially, in such cases, the RTP's target search behavior is similar to that of a standard passive Brownian particle with a larger (effective) diffusion coefficient that explains the shorter average TPT.
	On the other hand, when the average time spent in the active phases is much larger than the average time spent in the passive one, as it is the case for point B in Fig.~\ref{fig:4}, the RTP's behavior is more comparable to that of an ABP (compare Fig.~\ref{fig:5}b to Fig.~3b and~c in Ref.~\cite{Zanovello2021}) with reactive paths in which the particle spends time in surfing along the energy walls before falling into the target basin, thus giving rise to the long-time tails observed in Fig.~\ref{fig:3}.
	However, when a tumbling event occurs, the self-propulsion direction gets randomized and such randomization may allow the particle to stop surfing along the energy walls, thus enhancing, with respect to the completely active case, the relative frequency of reactive trajectory going from R to T by following a path close to the minimal energy path.
	The more frequent the tumbling events are, the more reactive paths follow this reactive channel, see Fig.~\ref{fig:5}c.
	Such an effect explains the peak at short times observed in Fig.~\ref{fig:3} and for case B and C in Fig.~\ref{fig:4}b.

	\section{\label{sec:conclusions} Conclusions}
	
	We have proposed a model for Run-and-Tumble dynamics of an agent alternating its behavior between passive and active Brownian motion.
	The two phases have exponentially distributed duration depending on a passive-to-active and an active-to-passive rate, respectively.
	Furthermore, a complete randomization of the self-propulsion direction occurs at the end of each passive phase.
	
	In line with what has already been elaborated for active Brownian particles (ABPs)~\cite{Zanovello2021}, we demonstrated here for the first time that we can extend the Transition-Path Sampling (TPS) algorithm to Run-and-Tumble particles (RTPs), thus virtually covering all relevant single-particle models of active matter.
	The key point, allowing such an extension of the TPS algorithm to active particles notwithstanding their lack of microscopic reversibility, is the definition of an appropriate backward dynamics and a consistent derivation of the acceptance probability that permits to sample the reactive-path ensemble in the Monte-Carlo-like scheme adopted by TPS.
	Our findings are corroborated by benchmarking the transition-path-time (TPT) distribution obtained by applying the TPS algorithm against that obtained by direct integration of the Langevin equations of motion.
	
	By exploiting this extension of the TPS algorithm, we investigated the target-search behavior of an RTP looking for a target that is separated from its starting position by an energy barrier. 
	More precisely, we characterized the transition-path ensemble in a paradigmatic energy landscape given by a double-well potential.
	Our findings show a rich phenomenology of the RTP's target-search conduct when changing the rates determining the average duration of the two phases.
	In particular, it results that tumbling is an efficient way of reducing the average transition-path time both with respect to a completely passive motion and to a completely active one.
	The main reason for this observation is that, while the activity of a particle is obviously helping in overcoming the energy barrier, the interplay of the persistence of the self-propulsion direction with the shape of the energy potential may instead extend the average TPT because long-lasting trajectories arise with the particle spending time surfing along the energy walls rather than meeting the target.
	On the other hand, the randomization of the self-propulsion direction occurring during tumbling events allows the RTP to detach more easily from the energy walls, thus enhancing the frequency of paths going relatively straight from the reactant to the target and its overall odds of reaching the target.
	Consequently, we also expect that for RPTs not only the average TPT is decreased, but also the overall rate of target finding events is increased when compared to an ABP having the same level of activity.
	However, here we didn't investigate the overall rate since this feature cannot be covered by the TPS algorithm alone if not complemented with transition interface sampling~\cite{verp2003,verp2005}, which goes beyond the scope of the present manuscript.
	
	The derivation of the TPS acceptance rule for new reactive paths of ABPs and RTPs may in principle be applied to other irreversible systems, such as chiral ABPs~\cite{vanTeeffelen2008}, active particles with anisotropic diffusion~\cite{Kurzthaler2016}, active Ornstein-Uhlenbeck particles~\cite{Bonilla2019}, and particles that orient themselves along spatial gradients~\cite{Pohl2014}.
	As a conclusion, we observe that our findinds contribute and may be relevant to further extend the general knowledge on how active particles perform in target-search problems~\cite{Schneider2019,Muinos-Landin2021,Zanovello2023,Caraglio2024} which is a crucial question in several realistic scenarios including bacteria foraging nourishment~\cite{Elgeti2015,Berg2004}, phagocytes of the immune system performing chemotactic motion during infection~\cite{Devreotes1988,Deoliveira2016}, and sperm cells on their way to the egg~\cite{Eisenbach2006}.
	In particular the target-search behavior of active particles in complex environments~\cite{Volpe2011} is central to many envisaged application of nanomaterials going from targeted drug delivery~\cite{Patra2013,Naahidi2013} to environmental remediation~\cite{Gao2014}.

\begin{acknowledgments}
	T.F. acknowledges funding by FWF: P 35580-N.~~M.C. is supported by FWF: P 35872-N.
\end{acknowledgments}
	\medskip

	\appendix
	\begin{widetext}

\section{\label{sec:fokkerplanck} Derivation of Fokker Planck equation}

	Here, we formulate the Fokker-Planck equation associated with the model presented in Eqs.~(\ref{eq:transitionprob_forward}-\ref{eq:eomrtpt_vartheta}).
	Usually, a well-defined procedure allowing to obtain the Fokker-Planck equations starting from the Langevin equations of motion is followed~\cite{Risken1989}.
	However, the presence of discontinuous updates in Eqs.~\eqref{eq:transitionprob_forward} and~\eqref{eq:eomrtpt_vartheta} makes it more convenient to follow a different approach.
	Namely, we start from the definition of the time-derivative of the propagator, we use the Chapman-Kolmogorov equation~\cite{Risken1989} to rephrase such a derivative in terms of single-step propagators and finally we insert the single-step propagator expression, as given in Eq.~\eqref{eq:p_singlestep_forward}, to derive the Fokker-Planck operator.

	The propagator $P_t(\omega| \omega_0)$ is the probability of going from the state $\omega_0$ to the state $\omega_1$ in a time $t$.
	Its temporal derivative can be written as
	\begin{align} \label{eq:A1}
		\frac{\partial}{\partial t} P_t(\omega| \omega_0) =\lim_{\Delta t\to 0} \frac{1}{\Delta t} \left\{ P_{t+\Delta t}(\omega|\omega_0) - P_t(\omega |\omega_0) \right\} =\lim_{\Delta t\to 0} \frac{1}{\Delta t}  \int \dd{\omega_1} \left[ \mathbb{P}_{\Delta t}(\omega|\omega_1) - \delta(\omega,\omega_1)   \right]   P_t(\omega_1 |\omega_0) \, ,
	\end{align}
	where we defined
	\begin{align}
		\int \dd{\omega} := \sum_{s=0}^{1} \, \int \dd{\vec{r}} \int_{-\infty}^{\infty} \dd{\vartheta_1}  \ ,
	\end{align}
	and
	\begin{align}
		\delta(\omega,\omega_1) := \delta_{s,s_1} \delta(\vec{r}-\vec{r}_1) \delta(\vartheta-\vartheta_1) \ ,
	\end{align}
	and where we used the Chapman-Kolmogorov equation with the propagator $\mathbb{P}_{\Delta t}(\omega|\omega_1)$ over a single step $\Delta t$
	\begin{align}
		P_{t+\Delta t}(\omega|\omega_0) = \int \dd{\omega_1}  \mathbb{P}_{\Delta t}(\omega|\omega_1)     P_t(\omega_1 |\omega_0) \, .
	\end{align}
	From Eq.~\eqref{eq:transitionprob_forward} the single-step propagator is
	\begin{align}  \label{eq:A5}
		\mathbb{P}_{\Delta t}(\omega|\omega_1) & =
			\frac{p_{\Delta t}(s |s_1)}{4\pi D \Delta t}  \exp\left( - \dfrac{\left( \vec{r}- \vec{r}_1 +\mu \vec{\nabla}_{\! \vec{r_1}} U(\vec{r}_1) \Delta t - v s_1  \vec{u}_1\Delta t \right)^2}{4 D \Delta t} \right) \left[ \frac{\delta_{s_1, 0}\delta_{s,1}}{2\pi}+ \frac{(1-\delta_{s_1, 0}\delta_{s,1})}{\sqrt{4\pi D_\vartheta \Delta t} } \exp\left( -\dfrac{(\vartheta-\vartheta_1)^2}{4 D_\vartheta \Delta t} \right) \right] \notag\\ 
			&= p_{\Delta t}(s| s_1) \int \dfrac{ \dd{\vec{k}} }{(2\pi)^3} \, \exp\left(- k^2 D \Delta t\right) \, \exp \left\{ i \vec{k} \cdot [ \vec{r} -\vec{r}_1  +\mu \vec{\nabla}_{\! \vec{r_1}} U(\vec{r}_1) \Delta t - v s_1  \vec{u}_1\Delta t ] \right\} \notag \\
			&  \quad \times  \left\{ \delta_{s_1, 0}\delta_{s,1}+ (1-\delta_{s_1, 0} \delta_{s,1}) \int_{-\infty}^\infty  \dd{\nu} \, \exp\left(-\nu^2D_\vartheta \Delta t \right) \, \exp \left[ i \nu(\vartheta-\vartheta_1)\right] \right\} \, ,
	\end{align}
	where we specified that the gradient of the potential is taken with respect to the variable $\vec{r_1}$, we used the Kronecker delta to account for the two possible cases of the angular update, and we exploited the Hubbard-Stratonovich transform~\cite{Hubbard1959} from the first to the second line.
	By Taylor expanding the exponential appearing in the previous formula we have
	\begin{align} \label{eq:A6}
		p_{\Delta t}(s |s_1) = \delta_{s,s_1}  + (1-2\delta_{s,s_1}) \lambda_{s_1 \rightarrow (1-s_1)} \Delta t + O(\Delta t^2) \, ,
	\end{align}
	\begin{align} \label{eq:A7}
		\frac{1}{(2\pi)^2} \int \dd{\vec{k}} &  \, \exp\left(- k^2 D \Delta t\right) \, \exp \left\{ i \vec{k} \cdot [ \vec{r} -\vec{r}_1  +\mu \vec{\nabla}_{\! \vec{r_1}} U(\vec{r}_1) \Delta t - v s_1  \vec{u}_1\Delta t ] \right\} \notag  \\
		& = \delta(\vec{r} -\vec{r}_1) - \frac{\Delta t}{(2\pi)^2}   \int \dd{\vec{k}} \, \left\{ k^2 D  - i \vec{k} \cdot [ \mu \vec{\nabla}_{\! \vec{r_1}} U(\vec{r}_1)  - v s_1  \vec{u}_1 ] \right\} \, e^{i \vec{k} \cdot (\vec{r} -\vec{r}_1)} + O(\Delta t^2) \, ,
	\end{align}
	\begin{align} \label{eq:A8}
		\frac{1}{(2\pi)} & \int_{-\infty}^\infty  \dd{\nu} \, \exp\left(-\nu^2D_\vartheta \Delta t \right) \, \exp \left[ i \nu(\vartheta-\vartheta_1)\right] = \delta(\vartheta-\vartheta_1) - \frac{\Delta t}{(2\pi)} \int_{-\infty}^\infty  \dd{\nu} \, \nu^2D_\vartheta \, e^{i \nu(\vartheta-\vartheta_1)}  + O(\Delta t^2) \, .
	\end{align}
	Inserting Eqs.~(\ref{eq:A6}-\ref{eq:A8}) into Eq.~\eqref{eq:A5}, some tedious but straightforward calculation lead us to
	\begin{align}  \label{eq:A9}
		\mathbb{P}_{\Delta t} (\omega|\omega_1) &  = \updelta (\omega - \omega_1) - \Delta t \,  \Bigg\{ (2\delta_{s,s_1}-1) \lambda_{s_1 \rightarrow (1-s_1)} \delta(\vec{r}-\vec{r_1}) \left[  \dfrac{\delta_{s_1, 0}\delta_{s,1}}{2\pi} + (1-\delta_{s_1, 0}\delta_{s,1} ) \delta(\vartheta-\vartheta_1) \right] \notag  \\
			 	& \qquad + \delta_{s,s_1} \delta(\vartheta-\vartheta_1) \int \dfrac{\dd{\vec{k}}}{(2\pi)^2} \, \left\{ k^2 D  - i \vec{k} \cdot [ \mu \vec{\nabla}_{\! \vec{r_1}} U(\vec{r}_1)  - v s_1  \vec{u}_1 ] \right\} \, e^{i \vec{k} \cdot (\vec{r} -\vec{r}_1)} \notag  \\
			 	& \qquad + \, \delta_{s,s_1}\; \delta(\vec{r}-\vec{r}_1) \; \int_{-\infty}^\infty  \dfrac{\dd{\nu}}{2\pi}  \, \nu^2D_\vartheta \, e^{i \nu(\vartheta-\vartheta_1)}  \Bigg\} + O(\Delta t^2) \; .
	\end{align}
	Finally, inserting the previous equation into~\eqref{eq:A1} results in
	\begin{align} \label{eq:A10}
		\frac{\partial}{\partial t} P_t(\omega| \omega_0) &=
		- \sum_{s_1=0}^{1} \, \int \dd{\vec{r}_1} \int_{-\infty}^{\infty} \dd{\vartheta_1}
		\Bigg\{ (2\delta_{s,s_1}-1) \lambda_{s_1 \rightarrow (1-s_1)} \delta(\vec{r}-\vec{r_1}) \left[  \dfrac{\delta_{s_1, 0}\delta_{s,1}}{2\pi} + (1-\delta_{s_1, 0}\delta_{s,1} ) \delta(\vartheta-\vartheta_1) \right] \notag  \\
			 	& \qquad + \delta_{s,s_1} \delta(\vartheta-\vartheta_1) \int \dfrac{\dd{\vec{k}}}{(2\pi)^2} \, \left\{ k^2 D  - i \vec{k} \cdot [ \mu \vec{\nabla}_{\! \vec{r_1}} U(\vec{r}_1)  - v s_1  \vec{u}_1 ] \right\} \, e^{i \vec{k} \cdot (\vec{r} -\vec{r}_1)} \notag  \\
			 	& \qquad + \, \delta_{s,s_1} \; \delta(\vec{r}-\vec{r}_1) \; \int_{-\infty}^\infty  \dfrac{\dd{\nu}}{2\pi} \, \nu^2D_\vartheta \, e^{i \nu(\vartheta-\vartheta_1)}   \Bigg\} P_t(\omega_1 |\omega_0) \; .
	\end{align}
	Integrating twice by parts with respect to $\vec{r_1}$ and to $\vartheta_1$ and imposing vanishing condition at $|r_1| \to \infty$ and $\vartheta_1 \to \pm \infty$ we have
	\begin{align} \label{eq:A11}
		& \frac{\partial}{\partial t}  P_t(\omega| \omega_0) =
		\sum_{s_1=0}^{1} \, \int \dd{\vec{r}_1} \int_{-\infty}^{\infty} \dd{\vartheta_1}
		\Bigg\{ (1-2\delta_{s,s_1}) \lambda_{s_1 \rightarrow (1-s_1)} \delta(\vec{r}-\vec{r_1}) \left[  \dfrac{\delta_{s_1, 0}\delta_{s,1}}{2\pi} + (1-\delta_{s_1, 0}\delta_{s,1} ) \delta(\vartheta-\vartheta_1) \right]  P_t(\omega_1 |\omega_0) \notag  \\
			 	& \quad + \delta(\omega,\omega_1) \Big[ D \nabla^2_{\! \vec{r_1}}   P_t(\omega_1 |\omega_0) + \mu  \vec{\nabla}_{\! \vec{r_1}}  \cdot [ P_t(\omega_1 |\omega_0) \vec{\nabla}_{\! \vec{r_1}} U(\vec{r}_1)  ] - v s_1  \vec{u}_1  \cdot \vec{\nabla}_{\! \vec{r_1}}  P_t(\omega_1 |\omega_0)  + \, D_\vartheta \, \partial_{\vartheta_1}^2  P_t(\omega_1 |\omega_0) \Big] \Bigg\} \; , 
	\end{align}
	which finally leads to
	\begin{align} \label{eq:A12}
		\frac{\partial}{\partial t}  P_t(\omega| \omega_0) = & \; (1-2\delta_{s,0}) \lambda_{0 \to 1} \int_{-\infty}^{\infty} \dd{\vartheta_1}    \left[  \dfrac{\delta_{s,1}}{2\pi} + (1-\delta_{s,1} ) \delta(\vartheta-\vartheta_1) \right]  P_t(0,\vec{r},\vartheta_1 |\omega_0) \notag  \\
		        & + (1-2\delta_{s,1}) \lambda_{1 \to 0}  P_t(1,\vec{r},\vartheta |\omega_0) \notag  \\
			 	& +   D \nabla^2   P_t(\omega |\omega_0) + \mu  \vec{\nabla}  \cdot [ P_t(\omega |\omega_0) \vec{\nabla} U(\vec{r})  ] - v s \vec{u}  \cdot \vec{\nabla}  P_t(\omega |\omega_0)  + \, D_\vartheta \, \partial_{\vartheta}^2  P_t(\omega |\omega_0) \; ,
	\end{align}
	where $\vec{\nabla} := \vec{\nabla}_{\! \vec{r}}$. Splitting over the two possible phases concludes the set of master equations 
	\begin{align} 
		\label{eq:A13}
		\partial_t P_t(0, \vec{r}, \vartheta |\omega_0) &= -\lambda_{0\to1} P_t(0,  \vec{r} ,\vartheta  |\omega_0) + \lambda_{1\to 0} P_t(1,  \vec{r} ,\vartheta  |\omega_0) + \mathcal{L}_0  P_t(0  ,\vec{r} ,\vartheta |\omega_0) \, , \\
		\label{eq:A14}
		\partial_t P_t(1,  \vec{r}, \vartheta |\omega_0) &= \lambda_{0\to1} \int \frac{\dd \vartheta_1}{2\pi}P_t(0, \vec{r},\vartheta_1 |\omega_0) - \lambda_{1\to 0} P_t(1,  \vec{r}, \vartheta |\omega_0) + (\mathcal{L}_0 + v \mathcal{L}_1) P_t(1,  \vec{r}, \vartheta |\omega_0) \, ,
	\end{align}
where we defined
\begin{align}
	&\mathcal{L}_0 := \vec{\nabla}  \cdot \left[ D \vec{\nabla} + \mu \vec{\nabla} U(\vec{r}) \right] +  D_\vartheta \partial_{\vartheta}^2 \, ,
	\notag \\
	&\mathcal{L}_1 := - \vec{u} \cdot \vec{\nabla} \, .
\end{align}

\section{Fokker-Planck Equation in the limit of the passive-to-active rate going to infinity ($\lambda_{0 \to 1}\rightarrow \infty$)}\label{appendix_infinityrate}

	We find a formal solution of the Fokker-Planck equation~\eqref{eq:A13}, by considering the ansatz
	\begin{align}  \label{eq:B1}
		P_t(0, \vec{r}, \vartheta |\omega_0) &=  e^{(\mathcal{L}_0 - \lambda_{0\to 1})t} c(\omega_0; t) \; ,
	\end{align}
	where $c(\omega_0; t)$ is an arbitrary function to be determined with the initial condition specified by taking $t=0$ in the previous equation and considering that the state cannot evolve in a time equal to zero
	\begin{align}  \label{eq:B2}
		c(\omega_0; 0) = \delta_{0,s_0} \delta(\vec{r}-\vec{r}_0) \delta(\vartheta-\vartheta_0)
	\end{align}
	Inserting into Eq.~\eqref{eq:A13} we have
	\begin{align}  \label{eq:B3}
		\partial_t c(\omega_0; t) =  \lambda_{1\to 0} \, e^{-(\mathcal{L}_0 - \lambda_{0\to 1})t} \, P_t(1,  \vec{r} ,\vartheta  |\omega_0) \; .
	\end{align}
	By integrating over time we find
	\begin{align}  \label{eq:B4}
		c(\omega_0; t) =  c(\omega_0; 0) + \lambda_{1\to 0} \int_0^t \dd{t'} \, e^{-(\mathcal{L}_0 - \lambda_{0\to 1})t'} \, P_{t'}(1,  \vec{r} ,\vartheta  |\omega_0) \; ,
	\end{align}
	which can then inserted back into the ansatz~\eqref{eq:B1} to obtain the expression for the propagator $P_t(0, \vec{r}, \vartheta |\omega_0)$
	\begin{align}  \label{eq:B5}
		P_t(0, \vec{r}, \vartheta |\omega_0) =  e^{(\mathcal{L}_0 - \lambda_{0\to 1})t}  \delta_{0,s_0} \delta(\vec{r}-\vec{r}_0) \delta(\vartheta-\vartheta_0) + \lambda_{1\to 0} \int_0^t \dd{t'} \, e^{(\mathcal{L}_0 - \lambda_{0\to 1})(t-t')} \, P_{t'}(1,  \vec{r} ,\vartheta  |\omega_0)  \; .
	\end{align}
	The first term on the right side of the previous equation can then be interpreted as the probability of evolving, while being in a passive phase, starting from an initial passive state having $\vec{r}=\vec{r}_0$ and $\vartheta=\vartheta_0$.
	On the other hand, the second term is the probability of evolving passively once the particle enters the passive phase at time $t'$, integrated over all possible times $t'$.
	In the limit of infinite passive-to-active rate going to infinity ($\lambda_{0 \to 1}\rightarrow \infty$) we then have for all $t>0$
	\begin{align}  \label{eq:B6}
		\lim_{\lambda_{0\to1}\to \infty} P_t(0, \vec{r}, \vartheta |\omega_0) =  0 \; .
	\end{align}

	Now we can insert Eq.~\eqref{eq:B4} into Eq.~\eqref{eq:A14} to obtain
	\begin{align}  \label{eq:B7}
		\partial_t P_t(1,  \vec{r}, \vartheta |\omega_0) = &\, \lambda_{0\to1} \int \frac{\dd \vartheta_1}{2\pi} \left\lbrace  
		e^{(\mathcal{L}_0 - \lambda_{0\to 1})t}  \delta_{0,s_0} \delta(\vec{r}-\vec{r}_0) \delta(\vartheta_1-\vartheta_0) + \lambda_{1\to 0} \int_0^t \dd{t'} \, e^{(\mathcal{L}_0 - \lambda_{0\to 1})(t-t')} \, P_{t'}(1,  \vec{r} ,\vartheta_1  |\omega_0)
		    \right\rbrace  \notag \\
		& - \lambda_{1\to 0} P_t(1,  \vec{r}, \vartheta |\omega_0) + (\mathcal{L}_0 + v \mathcal{L}_1) P_t(1,  \vec{r}, \vartheta |\omega_0) \, .
	\end{align}
	Taking the limit $\lambda_{0 \to 1}\rightarrow \infty$, the previous equation becomes
	\begin{align}  \label{eq:B8}
		\lim_{\lambda_{0\to1}\to \infty} \partial_t P_t(1,  \vec{r}, \vartheta |\omega_0) =   \lambda_{1\to 0} \int \frac{\dd \vartheta_1}{2\pi}    P_{t}(1,  \vec{r} ,\vartheta_1  |\omega_0) - \lambda_{1\to 0} P_t(1,  \vec{r}, \vartheta |\omega_0) + (\mathcal{L}_0 + v \mathcal{L}_1) P_t(1,  \vec{r}, \vartheta |\omega_0)\, ,
	\end{align}
	where we used the identity $\lim_{\lambda_{} \to \infty}\lambda_{} \exp(-\lambda_{}(t-t’))=\delta(t-t')$ and directly performed the integration over time $t'$.
	\medskip
	
	The former expression is indeed equivalent to the Fokker-Planck equation
	\begin{align}
		\partial_t P_t(1,  \vec{r}, \vartheta |\omega_0) =  \lambda \int \frac{\dd \vartheta_1}{2\pi}    P_{t}(1,  \vec{r} ,\vartheta_1  |\omega_0) - \lambda P_t(1,  \vec{r}, \vartheta |\omega_0) + (\mathcal{L}_0 + v \mathcal{L}_1) P_t(1,  \vec{r}, \vartheta |\omega_0)\, ,
	\end{align}
	that can be obtained following the same steps as in \cref{sec:fokkerplanck}, starting from the equation of motion of an ABP undergoing random updates of the self-propulsion direction with a rate $\lambda$
	\begin{align} 
	 \label{eq:B10}
	\vec{r}_{i+1} &= \vec{r}_{i}- \mu \vec{\nabla}U(\vec{r}_{i}) \Delta t+ \sqrt{2 D \Delta t } \boldsymbol{\xi}_{i}+  v \vec{u}_{i} \Delta t \\
	\label{eq:B11}
	{\vartheta}_{i+1} &= \begin{cases}
		\vartheta_{i} + \sqrt{2 D_\vartheta \Delta t} \eta_{i} &\text{with probability } \exp(-\lambda \Delta t)\\
		2\pi \, \zeta_{i} &\text{with probability } 1-\exp(-\lambda \Delta t)
	\end{cases}
\end{align}

	After performing the analytical limit, we compare the TPT distributions for RTP particles with instantaneous tumbling events, as obtained by directly integrating Eqs.~\eqref{eq:B10},~\eqref{eq:B11}, to the one obtained in our model with infinite passive-to-active rate, $\lambda_{0\to1}\to \infty$. 
	As expected, the two models yield the same TPT distribution, as long as the integration time step is small enough, see Fig.~\ref{fig:6} .

\begin{figure*}[t!] 
	\centering
	\includegraphics[width=1.0\textwidth]{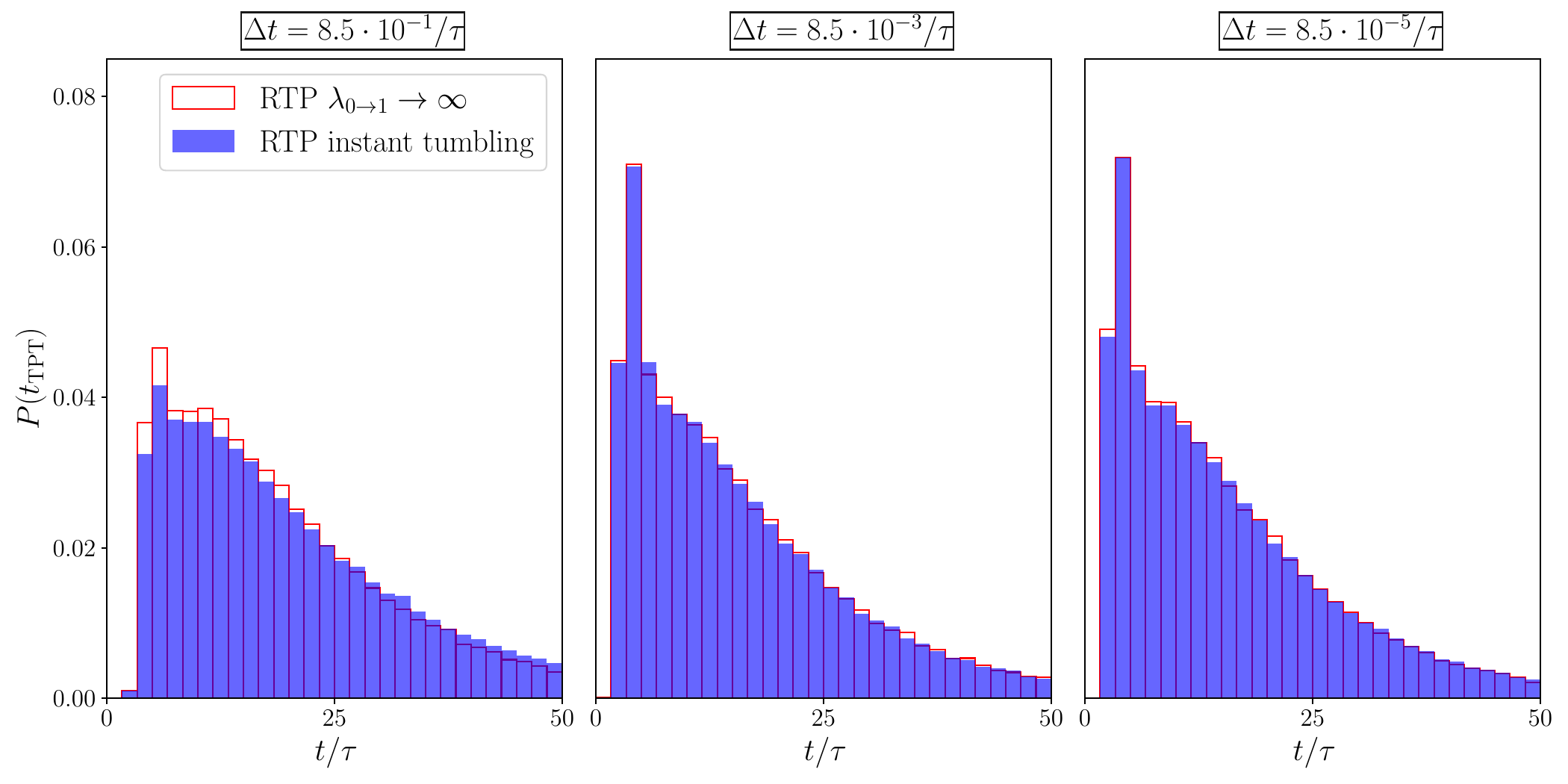}
	\caption{Transition-Path-time distribution of an RTP with infinite passive-to-active rate and an active-to-passive rate of $\lambda_{1\to 0}=0.12/\tau$ (empty bars) compared to the distribution obtained with a Run-and-Tumble model with instantaneous reorientation between two consecutive runs and a tumble rate of $\lambda = 0.12/\tau$ (full bars). \label{fig:6}
}	
\end{figure*}

\section{ \label{sec:pathint}Stochastic Path Integral for Run-and-Tumble particles}

	Starting from an initial distribution $P_0(\omega_0; t_0)$ at time $t_0=0$, the single step propagator $\mathbb{P}_{\! \Delta t}$ as given in Eq.~\eqref{eq:P(W)_general} allows us to express the probability of being in state $\omega_N$ at time $t=N \Delta t$ as
	\begin{align}\label{eq:C1}
		P(\omega_N;N\Delta t)   =  \int \dd{\omega_0} \dd{\omega_1} \ldots  \dd{\omega_{N-1}} \left[  \prod_{i=0}^{N-1} \mathbb{P}_{\! \Delta t}(\omega_{i+1}| \omega_{i}) \, P(\omega_0;0) \right] \; .
	\end{align}
	
	We now define a switching step a time step during which it occurs a transition from the state $\omega_i=(s_i,\vec{r}_i,\vartheta_i)$ to the state $\omega_{i+1}=(s_{i+1},\vec{r}_{i+1},\vartheta_{i+1})$ such that $s_{i+1} = 1-s_i$.
	By defining $\mathcal{W}^{0 \to 1}_{i,i+1} := ((0,\vec{r}_i,\vartheta_i),(1,\vec{r}_{i+1},\vartheta_{i+1}))$ and $\mathcal{W}^{1 \to 0}_{i,i+1} := ((1,\vec{r}_i,\vartheta_i),(0,\vec{r}_{i+1},\vartheta_{i+1}))$, the probabilities of such transitions read 
	\begin{align}\label{eq:C2}
		\mathcal{P} (\mathcal{W}^{0 \to 1}_{i,i+1}) =  \mathbb{P}_{\Delta t} (1 , \vec{r}_{i+1}, \vartheta_{i+1}|0 , \vec{r}_{i}, \vartheta_{i}) = \dfrac{ (1-\exp{-\lambda_{0 \to 1} \Delta t})}{2\pi (4\pi D \Delta t)} \exp \left\{ - \frac{\left[ \vec{r}_{i+1}- \vec{r}_{i} +\mu \vec{\nabla}U (\vec{r}_{i}) \Delta t\right]^2}{4 D \Delta t} \right\} \; ,
	\end{align}
	and 
	\begin{align}\label{eq:C3}
		\mathcal{P} (\mathcal{W}^{1 \to 0}_{i,i+1}) & =  \mathbb{P}_{\Delta t} (0 , \vec{r}_{i+1}, \vartheta_{i+1}|1 , \vec{r}_{i}, \vartheta_{i}) \notag \\
		& = \dfrac{ (1-\exp{-\lambda_{1 \to 0} \Delta t})}{ (4\pi D \Delta t) (4\pi D_{\vartheta} \Delta t)^{1/2}} \exp \left\{ - \frac{\left[ \vec{r}_{i+1}- \vec{r}_{i} +\mu \vec{\nabla}U (\vec{r}_{i}) \Delta t - v \vec{u}_i \Delta t \right]^2}{4 D \Delta t} \right\}  \exp \left\{ - \frac{( \vartheta_{i+1}- \vartheta_{i})^2}{4 D_{\vartheta} \Delta t} \right\} \; ,
	\end{align}
	respectively, see Eq.~\eqref{eq:p_singlestep_forward}.
	Given a particular trajectory $\mathcal{W} = (\omega_0,\omega_1, \ldots, \omega_N)$, we can identify $M$ ($0 \leq M \leq N$) phase-switching steps occuring at the indexes $\mathcal{J}_M = (j_1,j_2,\ldots,j_M) = (i | 0\leq i < N \mbox{ and } s_{i+1} = 1-s_i)$ and corresponding to the times $(\tau_1 = j_1\Delta t, \tau_2 = j_2\Delta t, \ldots , \tau_M = j_M\Delta t)$.
	Note that if $s_N=s_0$ then $M$ has to be even, while if $s_N=1-s_0$ then $M$ has to be odd.
	In between phase-switching steps, we either have a continuous sequence of passive phases or a continuous sequence of active ones.
	The transition probability of a passive stretch $\mathcal{W}^{0}_{i,k} := ((0,\vec{r}_i,\vartheta_i),(0,\vec{r}_{i+1},\vartheta_{i+1}), \ldots, (0,\vec{r}_k,\vartheta_k))$ is
	\begin{align}\label{eq:C4}
		\mathcal{P} (\mathcal{W}^{0}_{i,k}) & =   \delta_{s_k,0} \prod_{l=i}^{k-1} \delta_{s_l,0} \, \mathbb{P}_{\! \Delta t}(\omega_{l+1}| \omega_{l}) \notag  \\
		& = 
		\dfrac{ \exp{-(k-i)\lambda_{0 \to 1} \Delta t} }{(4\pi D \Delta t)^{k-i} \, (4\pi D_\vartheta \Delta t)^{(k-i)/2}   } 
	\prod_{l=i}^{k-1} \exp \left\{ - \frac{\left[ \vec{r}_{l+1}- \vec{r}_{l} +\mu \vec{\nabla}U (\vec{r}_{l}) \Delta t \right]^2}{4 D \Delta t} \right\} \exp \left\{ - \frac{\left( \vartheta_{l+1} - \vartheta_{l} \right)^2}{4 D_\vartheta \Delta t} \right\} \, .
	\end{align}
	Analogously, an active stretch $\mathcal{W}^{1}_{i,k} := ((1,\vec{r}_i,\vartheta_i),(1,\vec{r}_{i+1},\vartheta_{i+1}), \ldots, (1,\vec{r}_k,\vartheta_k))$ has a transition probability
	\begin{align}\label{eq:C5}
		\mathcal{P} (\mathcal{W}^{1}_{i,k})  & =   \delta_{s_k,1} \prod_{l=i}^{k-1} \delta_{s_l,1} \, \mathbb{P}_{\! \Delta t}(\omega_{l+1}| \omega_{l}) \notag  \\
		& = 
		\dfrac{ \exp{-(k-i)\lambda_{1 \to 0} \Delta t} }{(4\pi D \Delta t)^{k-i} \, (4\pi D_\vartheta \Delta t)^{(k-i)/2}   } 
	\prod_{l=i}^{k-1} \exp \left\{ - \frac{\left[ \vec{r}_{l+1}- \vec{r}_{l} +\mu \vec{\nabla}U (\vec{r}_{l}) \Delta t - v \vec{u}_l \Delta t \right]^2}{4 D \Delta t} \right\} \exp \left\{ - \frac{\left( \vartheta_{l+1} - \vartheta_{l} \right)^2}{4 D_\vartheta \Delta t} \right\} \, .
	\end{align}
	
	Equation~\eqref{eq:C1} can then be rewritten as
	\begin{align}\label{eq:C6}
		P(\omega_N;N\Delta t)   & =  \sum_{s_0=0}^1 \sum_{M=0}^N \delta_{|s_N-s_0|,\text{mod}(M,2)} \sum_{\{\mathcal{J}_M\}} \int \dd{\vec{r}_0} \dd{\vec{r}_1} \ldots \ \dd{\vec{r}_{N-1}} \, \dd{\vartheta_0}  \dd{\vartheta_1}  \ldots \ \dd{\vartheta_{N-1}} \notag \\
		& \quad \left[  \left( \ldots         
		      \mathcal{P} (\mathcal{W}^{0}_{j_i+1,j_{i+1}})    
		      \mathcal{P} (\mathcal{W}^{1 \to 0}_{j_i,j_i+1})
		      \mathcal{P} (\mathcal{W}^{1}_{j_{i-1}+1,j_i})
		      \mathcal{P} (\mathcal{W}^{0 \to 1}_{j_{i-1},j_{i-1}+1})
		      \mathcal{P} (\mathcal{W}^{0}_{j_{i-2}+1,j_{i-1}})
		  \ldots \right)  \, P(\omega_0;0) \right] \; ,
	\end{align}
	where the summation over all the phase variables $\sum_{s_0=0}^1 \sum_{s_1=0}^1 \ldots \sum_{s_{N-1}=0}^1$ has been substituted by a summation over the first phase only $\sum_{s_0=0}^1$, a summation over the allowed numbers $M$ of phase switches $\sum_{M=0}^N \delta_{|s_N-s_0|,\text{mod}(M,2)}$, and a summation over all possible realization of $M$ phase switches $\sum_{\{\mathcal{J}_M\}}$.
	Furthermore, the exact sequence of probabilities $\mathcal{P}$ appearing in the second line of Eq.~\eqref{eq:C6} depends on the given value of $s_N$ and on the one of $s_0$ selected by the first summation.
	
	For infinitesimal time steps we can then symbolically write
	\begin{align}\label{eq:C7}
		P(\omega; t)  & =  \dfrac{1}{\mathcal{Z}}   \int \mathcal{D} s \, \, \mathcal{D} \vec{r} \, \mathcal{D} \vartheta
 \Bigg[  \Bigg( \ldots          
		      \dfrac{\lambda_{0 \to 1} e^{-\lambda_{0 \to 1}(\tau_{i+1}-\tau_{i}) }}{2\pi} e^{-S_0[\vec{r},\vartheta; \tau_{i},\tau_{i+1}]} \notag \\  
		      & \qquad \lambda_{1 \to 0} e^{-\lambda_{1 \to 0}(\tau_{i}-\tau_{i-1}) } e^{-S_1[\vec{r},\vartheta; \tau_{i-1},\tau_{i}]}    \notag \\
		      & \qquad \dfrac{\lambda_{0 \to 1} e^{-\lambda_{0 \to 1}(\tau_{i-1}-\tau_{i-2}) }}{2\pi} e^{-S_0[\vec{r},\vartheta; \tau_{i-2},\tau_{i-1}]}
		  \ldots \Bigg)  \, P(\omega_0;0) \Bigg] \; ,
	\end{align}
	with $\mathcal{Z}$ a normalization constant and
	\begin{align}
		S_0[\vec{r},\vartheta; \tau_{1},\tau_{2}] := \int_{\tau_1}^{\tau_2} \dd{t} \left\lbrace 
		\dfrac{\left[\vec{\dot{r}}(t)+\mu\vec{\nabla}U(\vec{r}(t))\right]^2}{4D} + \dfrac{ \left[ \dot{\vartheta}(t) \right]^2 }{4D_\vartheta}
		\right\rbrace \; ,
	\end{align}
	\begin{align}
		S_1[\vec{r},\vartheta; \tau_{1},\tau_{2}] := \int_{\tau_1}^{\tau_2} \dd{t} \left\lbrace 
		\dfrac{\left[\vec{\dot{r}}(t)+\mu\vec{\nabla}U(\vec{r}(t)) -v \vec{u}(t)\right]^2}{4D} + \dfrac{ \left[ \dot{\vartheta}(t) \right]^2 }{4D_\vartheta}
		\right\rbrace \; ,
	\end{align}
	and where we used the fact that
	\begin{align}\label{eq:C8}
	\left( \prod_{l=i}^{k-1} e^{-\lambda \Delta t} \right) \left( 1- e^{-\lambda \Delta t}\right) = e^{-\lambda (k-i) \Delta t}  -  e^{-\lambda (k-i+1) \Delta t} \xrightarrow[\Delta t \to 0]{} -\dd{t} \, \dfrac{\partial}{\partial t} e^{-\lambda t}  = \lambda e^{-\lambda t} \dd{t}\; ,
	\end{align}
	with the $\dd{t}$ then absorbed in the symbolic expression $ (\mathcal{D} s \, \, \mathcal{D} \vec{r} \, \mathcal{D} \vartheta)$ appearing in the path integral formulation and $p(t) = \lambda e^{-\lambda t}$ the PDF of an exponential distribution with parameter $\lambda$.
	We also used the fact that the remaining part of the switching probabilities~\eqref{eq:C2} and~\eqref{eq:C3} reduce to Dirac deltas in the limit of $\Delta t \to 0$
	\begin{align}\label{eq:C9}
	\lim_{\Delta t \to 0} \dfrac{ 1}{2\pi (4\pi D \Delta t)} \exp \left\{ - \frac{\left[ \vec{r}_{i+1}- \vec{r}_{i} +\mu \vec{\nabla}U (\vec{r}_{i}) \Delta t\right]^2}{4 D \Delta t} \right\} = \dfrac{ \delta(\vec{r}_{i+1}-\vec{r}_i)}{2 \pi} \; ,
	\end{align}
	and 
	\begin{align}\label{eq:C10}
		\lim_{\Delta t \to 0}  \dfrac{ 1}{ (4\pi D \Delta t) (4\pi D_{\vartheta} \Delta t)^{1/2}} & \exp \left\{ - \frac{\left[ \vec{r}_{i+1}- \vec{r}_{i} +\mu \vec{\nabla}U (\vec{r}_{i}) \Delta t - v \vec{u}_i \Delta t \right]^2}{4 D \Delta t} \right\}  \exp \left\{ - \frac{( \vartheta_{i+1}- \vartheta_{i})^2}{4 D_{\vartheta} \Delta t} \right\} \notag \\
		& =   \delta(\vec{r}_{i+1}-\vec{r}_i) \delta(\vartheta_{i+1}-\vartheta_i)\; ,
	\end{align}
	respectively.
	
	\begin{figure*}[t!] 
	\centering
	\includegraphics[width=1.0\textwidth]{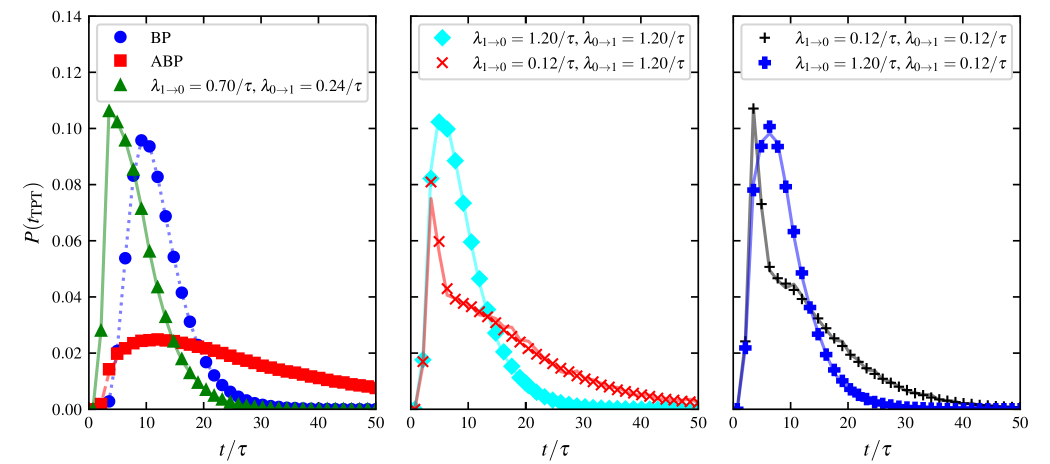}
	\caption{Transition-Path-Time distribution $P(t_{\rm TPT})$ generated with the Transition-Path Sampling (TPS) (lines) and compared to that obtained from standard MD simulations (symbols) for the pairs of parameters $\lambda_{1\to0}$ and $\lambda_{0\to1}$ reported in Fig.~\ref{fig:4}. Panel (a) also reports the TPT distribution of a fully passive Brownian particle -BP- and that of an ABP having the same $\text{Pe}$ and $\ell^*$ as the RTP. Statistics collected from $10^6$ reactive paths for standard MD simulations and $10^8$ reactive paths for TPS.
 \label{fig:7}
}	
\end{figure*}	
	
	\section{ \label{sec:appD} Additional comparison between TPT distributions obtained from the TPS and from standard MD simulations}
	
	For the combination of the parameters $\lambda_{0\to1}$ and $\lambda_{1\to 0}$ highlighed by the circles A-G in Fig.~\ref{fig:4}a, Fig.~\ref{fig:7} reports additional comparison between the distribution obtained by exploiting the TPS algorithm and those deriving from standard MD simulations.

\end{widetext}

\end{document}